\documentclass[11pt]{article}

\usepackage[margin=1in]{geometry}
\usepackage[T1]{fontenc}
\usepackage{microtype}
\usepackage{array}
\usepackage{booktabs}
\usepackage{graphicx}
\usepackage{multirow}
\usepackage{tabularx}
\usepackage{subfig}
\usepackage{amsmath}
\usepackage{amssymb}
\usepackage{authblk}
\usepackage{cite}
\usepackage[hidelinks]{hyperref}

\title{Tracking Harmfulness--Refusal Coupling under Dynamic Adversarial Fine-Tuning}

\author[1]{Wenhao~Lan}
\author[2]{Shan~Li}
\author[1]{Xinhua~Lai}
\author[3]{Meiqi~Wu}
\author[1]{Junbin~Yang}
\author[1]{Haihua~Shen\thanks{Corresponding author: \texttt{shenhh@ucas.ac.cn}}}
\affil[1]{University of Chinese Academy of Sciences, Beijing, China}
\affil[2]{Inner Mongolia University of Technology, Inner Mongolia, China}
\affil[3]{Tsinghua University, Beijing, China}
\date{}

\newcommand{\hrcirepr}{\ensuremath{\mathrm{HRCI}_{\mathrm{repr}}}}

\newcommand{\ind}{\mathbb{I}}
\newcommand{\keywords}[1]{\par\noindent\textbf{Keywords:} #1\par}

\begin{document}
\maketitle

\begin{abstract}
Safety alignment requires two linked capabilities: recognizing harmful requests and activating refusal when that recognition matters. Behavioral metrics alone cannot tell whether adversarial fine-tuning strengthens either capability or merely changes how tightly they are coupled. We study this problem with a dual safety-geometry protocol that extracts harmfulness carriers, refusal carriers, and direct harmfulness--refusal (H/R) representational coupling across aligned instruction-tuned anchors and matched Mistral-7B-v0.1 supervised fine-tuning (SFT) and Robust Refusal Dynamic Defense (R2D2) trajectories. The aligned anchors calibrate the protocol: refusal-side interventions reopen HarmBench attack success more strongly than harmfulness-only interventions, while the two carriers remain nearly orthogonal. The main result is a coupling-regime transition in R2D2. The primary direct-coupling metric, the representational Harmfulness--Refusal Coupling Index (\hrcirepr{}), drops from 0.0784 at step 50 to 0.0205 at step 500, while fixed-source attack success rate (ASR) rises from 0 to 0.25, XSTest refusal falls from 1.00 to 0.228, and benign helpfulness recovers from 0 to 1.12 on a 0--2 scale. SFT supplies the negative control: its direct-coupling change is much smaller, and it remains high-ASR, so low coupling is not itself a safety score. Causal sweeps, dense all-anchor checks, and sparse Greedy Coordinate Gradient (GCG) and AutoDAN transfer comparisons against refusal-drift predictors support H/R coupling as a fixed-protocol descriptive diagnostic rather than a standalone safety predictor. The evidence does not prove independent harmfulness and refusal pathways or deployment-grade prediction.
\end{abstract}

\section{Introduction}

A useful safety policy for a language model is selective. It should refuse requests for harmful assistance, but it should not refuse every prompt that contains sensitive terms, hypothetical framing, or safety-adjacent language. This selectivity depends on at least two internal functions: harmfulness recognition and refusal control. Attack success rates, refusal rates, and utility scores reveal the resulting behavior, but they do not identify which internal function changed. A model may look robust because it recognizes harmfulness more reliably, because its refusal policy fires too broadly, or because harmfulness recognition and refusal have become tightly coupled.

This ambiguity matters during adversarial fine-tuning. HarmBench, StrongREJECT, and XSTest make it possible to observe robust refusal, harmful usefulness, and over-refusal side by side \cite{mazeika2024harmbench,souly2024strongreject,rottger2024xstest}. Mechanistic work shows that refusal can often be represented by directions or subspaces in activation space \cite{arditi2024refusal,siu2025cosmic,wollschlager2025geometry}. Other work shows that harmfulness and refusal can be encoded separately \cite{zhao2025harmfulnessrefusal}. What remains unclear is how adversarial fine-tuning reorganizes the coupling between harmfulness recognition and refusal control over training, and whether that reorganization explains simultaneous changes in jailbreak robustness, over-refusal, and benign utility.

We study this question with a dual safety-geometry protocol. For each checkpoint, the protocol extracts a harmfulness carrier and a refusal carrier, compares their direct activation-space overlap, and separately reports their layer locations. We first calibrate the protocol on aligned Llama-3.1-8B-Instruct and Qwen2.5-7B-Instruct anchors. We then apply it to matched Mistral-7B-v0.1 standard supervised fine-tuning (SFT) and Robust Refusal Dynamic Defense (R2D2) trajectories, using fixed-source HarmBench Greedy Coordinate Gradient (GCG) attack success rate (ASR), XSTest refusal, StrongREJECT, benign utility, causal harmfulness--refusal (H/R) interventions, and sparse GCG/AutoDAN transfer as behavior readouts \cite{jiang2023mistral,grattafiori2024llama3,yang2024qwen25,mazeika2024harmbench,zou2023universal,liu2024autodan}.

Because a related refusal-geometry manuscript by overlapping authors analyzes part of the same Mistral SFT/R2D2 trajectory, we state the distinction explicitly \cite{lan2026refusalgeometry}. That manuscript treats refusal control as the primary internal object and studies KL-admissible refusal carriers along a robustness--utility frontier. Here the refusal carrier is only one side of a dual measurement system. The unit of analysis is H/R coupling: whether harmfulness recognition and refusal control become more or less coupled during dynamic adversarial fine-tuning. This change in unit requires new harmfulness-carrier extraction, direct H/R coupling diagnostics, aligned-anchor calibration, H/R-specific causal matrices, dense HRCI checks, and sparse-transfer comparisons against refusal-drift and behavior-state predictors.

\begin{figure*}[!t]
\centering
\includegraphics[width=0.98\textwidth]{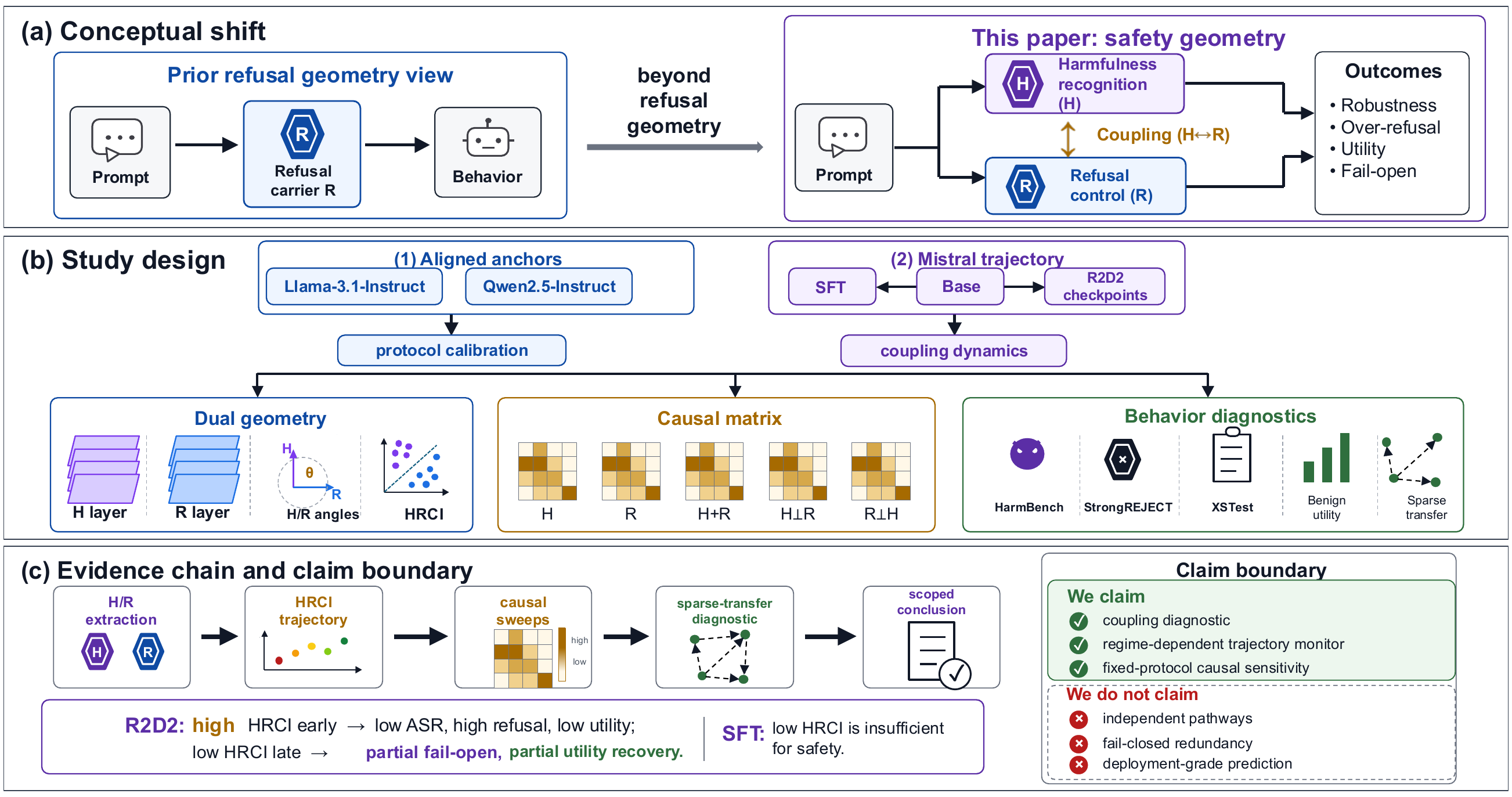}
\caption{From refusal geometry to harmfulness--refusal safety geometry. Panel (a) contrasts the refusal-carrier view studied in the related refusal-geometry manuscript \cite{lan2026refusalgeometry} with the dual H/R view studied here. Panel (b) shows the design: aligned instruction-tuned anchors calibrate the protocol, while matched Mistral SFT and R2D2 trajectories test how coupling changes under dynamic adversarial fine-tuning. Panel (c) gives the claim boundary. H/R coupling is used as an operational diagnostic, not as proof of independent pathways, fail-closed redundancy, or deployment-grade prediction.}
\label{fig:intro_overview}
\end{figure*}

The central finding is that R2D2 crosses two coupling regimes. Early R2D2 checkpoints have relatively higher direct H/R representational coupling, near-zero fixed-source ASR, saturated XSTest refusal, and collapsed benign helpfulness. Later checkpoints have much lower direct coupling, partial utility recovery, and reopened attack success. In the five-anchor summary, \hrcirepr{} falls from 0.0784 at step 50 to 0.0205 at step 500, while fixed-source ASR rises from 0 to 0.25 and benign helpfulness rises from 0 to 1.12 on a 0--2 scale. SFT prevents the simple interpretation that lower coupling means greater safety: its \hrcirepr{} changes only weakly, yet it remains high-ASR.

The paper makes three contributions:
\begin{enumerate}
    \item It gives a calibrated measurement protocol for tracking harmfulness carriers, refusal carriers, and their coupling across training.
    \item It identifies a regime transition in R2D2 that links internal H/R coupling to a robustness--over-refusal--utility transition, with SFT as a contrasting trajectory.
    \item It tests the boundary of the diagnostic with causal interventions, quality-controlled all-anchor descriptive checks, and sparse-transfer Harmfulness--Refusal Coupling Index (HRCI)-vs-drift comparisons.
\end{enumerate}
The result is not a claim that harmfulness and refusal are independent pathways. It is a narrower claim: coupling measurements help explain training states that output metrics alone conflate.

\section{Related Work}

\paragraph{Refusal directions, subspaces, and harmfulness--refusal separation.}
Refusal can often be localized in activation space. Arditi et al. show that refusal behavior in aligned models can be mediated by a single direction \cite{arditi2024refusal}; COSMIC identifies refusal directions with activation-space criteria rather than output templates \cite{siu2025cosmic}; and concept-cone work argues that refusal may occupy structured subspaces rather than a single vector \cite{wollschlager2025geometry}. These studies make targeted refusal interventions possible, but they do not by themselves show how harmfulness recognition relates to refusal control during training. The closest motivation for our protocol is the finding that large language models (LLMs) can encode harmfulness and refusal as distinct latent concepts \cite{zhao2025harmfulnessrefusal}. We use that separation as a starting point and ask how the relation between the two changes along SFT and R2D2 trajectories.

\paragraph{Safety evaluation under robustness, over-refusal, and utility tradeoffs.}
Automated red teaming and jailbreak benchmarks supply the behavioral axes for this study. HarmBench standardizes robust-refusal evaluation and introduces R2D2 \cite{mazeika2024harmbench}; GCG and AutoDAN provide fixed attack families for white-box and stealthier jailbreak stress tests \cite{zou2023universal,liu2024autodan}; and JailbreakBench emphasizes threat-model clarity \cite{chao2024jailbreakbench}. XSTest, the Over-Refusal Benchmark (OR-Bench), and FalseReject measure over-refusal or contextual false refusal \cite{rottger2024xstest,cui2024orbench,zhang2025falsereject}, while StrongREJECT distinguishes empty jailbreaks from harmful usefulness \cite{souly2024strongreject}. The MUBENCH benchmark makes the same safety--over-safety--utility tension explicit in machine unlearning for LLM safety \cite{zhao2026mubench}. We use these benchmarks as readouts for representation geometry: the target is not another behavior-only leaderboard, but an explanation of which internal relation changes when robustness, over-refusal, and utility move together.

\paragraph{Post-training safety dynamics.}
Fine-tuning can compromise safety even when the data are not intentionally malicious \cite{qi2024fine}. Defense-oriented work studies refusal-feature adversarial training (ReFAT), latent adversarial training, and projection constraints on refusal directions \cite{yu2025refat,casper2024latentAdversarial,sheshadri2024targetedLat,du2025procon}. A related refusal-geometry manuscript by overlapping authors analyzed part of the same Mistral SFT/R2D2 trajectory from a refusal-only perspective, asking whether dynamic adversarial fine-tuning reorganizes an admissible refusal-control carrier and how that carrier relates to a robustness--utility frontier \cite{lan2026refusalgeometry}. That study did not measure harmfulness carriers, direct harmfulness--refusal coupling, aligned-anchor H/R calibration, or H/R-specific causal matrices. This study uses that trajectory result as a motivating observation, but changes the research question and the measurement target: we ask whether refusal-geometry reorganization corresponds to a change in the coupling between harmfulness recognition and refusal control. The resulting analysis introduces aligned-anchor calibration, dual H/R carrier extraction, direct coupling diagnostics, orthogonalized H/R interventions, dense all-anchor HRCI checks, and sparse-transfer comparisons against refusal-drift and behavior-state predictors.

\section{Dual Safety-Geometry Protocol}
\label{sec:protocol}

The protocol has five steps. First, we fix probe sets, readout positions, candidate layers, and checkpoint anchors so that geometry quantities are comparable across training. Second, we extract harmfulness carriers from harmful-versus-benign contrasts and refusal carriers from refusal-versus-compliance contrasts. Third, we summarize direct H/R coupling using directional and subspace overlap, while reporting layer separation separately. Fourth, we test functional sensitivity with H-only, R-only, joint, orthogonalized, and matched-control interventions. Fifth, we align these internal quantities with fixed-source ASR, XSTest refusal, StrongREJECT, benign utility, dense all-anchor diagnostics, and sparse-transfer readouts. Throughout, the quantities are treated as fixed-protocol diagnostics rather than as standalone safety scores.

\subsection{Models, checkpoints, and behavioral axes}

The main trajectory uses Mistral-7B-v0.1 under two matched post-training regimes: standard supervised fine-tuning (SFT) and R2D2-style dynamic adversarial fine-tuning \cite{jiang2023mistral, mazeika2024harmbench}. Unless stated otherwise, both regimes are evaluated at five anchors: reference, step 50, step 100, step 250, and step 500. The reference is selected separately within each regime using a behavior-based rule: we take the earliest checkpoint at which direct-refusal behavior on the fixed probe set is stable. This yields step 30 for R2D2 and step 5 for SFT. Thus, ``reference'' denotes a regime-specific aligned anchor rather than the pretrained backbone or a shared optimization step. The aligned-anchor calibration uses instruction-tuned Llama-3.1-8B and Qwen2.5-7B models \cite{grattafiori2024llama3,yang2024qwen25}. These anchors test whether the H/R extraction and intervention protocol produces sensible behavior in models with mature instruction-following and safety behavior, under a dual-carrier framing motivated by harmfulness--refusal separation work \cite{zhao2025harmfulnessrefusal}.

\paragraph{Relationship to the shared trajectory artifact.}
The Mistral SFT/R2D2 checkpoints and several behavior readouts overlap with the trajectory analyzed in a related refusal-geometry manuscript by overlapping authors \cite{lan2026refusalgeometry}. In that manuscript, the checkpoints were used to study admissible refusal-control carriers, carrier relocation, effective rank, principal-angle refusal drift, sparse adaptive stress tests, and a robustness--utility frontier. We use the same trajectory as a controlled substrate for a different diagnostic question. All harmfulness-carrier extractions, direct H/R coupling quantities, aligned-anchor calibration runs, H/R causal matrices, dense all-anchor HRCI sensitivity analyses, and HRCI-vs-drift sparse-transfer comparisons are specific to this manuscript.

Behavior is measured on five axes. Fixed-source HarmBench GCG ASR measures attack success against a fixed harmful-behavior set \cite{mazeika2024harmbench,zou2023universal}. XSTest any-refusal measures over-refusal on safe prompts \cite{rottger2024xstest}. StrongREJECT score measures harmful usefulness rather than merely non-refusal \cite{souly2024strongreject}. Benign utility is measured on a 60-prompt continuity set with 0--2 helpfulness scoring, refusal flags, and degeneration flags, following the safety--over-safety--utility framing used in recent LLM safety evaluation work \cite{zhao2026mubench}. Sparse-transfer ASR evaluates checkpoint-specific GCG and AutoDAN attacks transferred across source and target checkpoints \cite{zou2023universal,liu2024autodan}.

All carrier extractions use fixed probe sets across checkpoints, fixed post-instruction readout positions, and a fixed candidate-layer grid within each protocol block. Layer indices are zero-based, and the hidden-state carriers are residual-stream contrast directions after L2 normalization. The subspace term uses the leading local H/R subspaces selected by the same extraction procedure as the carrier directions. Exact prompt-set sizes, split definitions, token positions, candidate layers, hook locations, subspace dimension, attack budgets, decoding parameters, judge settings, checksums, and run manifests are recorded in the supplementary artifact manifest and the source tables under the supplementary identifiers S1--S7 and S9.

All reported trajectory summaries use the single completed Mistral training run for each regime, so checkpoint-to-checkpoint trends are treated as ordered trajectory diagnostics rather than cross-seed estimates. The dense all-anchor intervals use the moving-block bootstrap specified in the source tables; sparse transfer is a fixed GCG/AutoDAN transfer readout rather than an adaptive or black-box attacker-model search.

\subsection{Notation and measured carriers}

Let \(M_t\) denote a checkpoint at anchor \(t\) within a fixed training regime, and let
\begin{equation}
    a^{(\ell)}_{p}(x;M_t)\in\mathbb{R}^{d}
\end{equation}
denote the residual-stream activation for input \(x\), layer \(\ell\), and token position \(p\).  The protocol selects two measurement carriers,
\begin{equation}
    h_t\in\mathbb{S}^{d-1},\qquad r_t\in\mathbb{S}^{d-1},
\end{equation}
where \(\mathbb{S}^{d-1}\) is the unit sphere in the \(d\)-dimensional hidden-state space.  Thus both carriers are L2-normalized, \(\|h_t\|_2=\|r_t\|_2=1\).  The harmfulness carrier \(h_t\) is obtained from harmful-versus-benign input contrasts.  Given harmful and benign prompt sets \(\mathcal{D}_{\mathrm{harm}}\) and \(\mathcal{D}_{\mathrm{benign}}\), we compute, at the harmfulness readout position \(p_H(t)\), the layerwise difference in means
\begin{equation}
    \Delta^H_{\ell,t}
    =
    \mathbb{E}_{x\in\mathcal{D}_{\mathrm{harm}}}
    a^{(\ell)}_{p_H(t)}(x;M_t)
    -
    \mathbb{E}_{x\in\mathcal{D}_{\mathrm{benign}}}
    a^{(\ell)}_{p_H(t)}(x;M_t).
    \label{eq:harmfulness_carrier_delta}
\end{equation}
A train/validation split is used to choose the most stable layer with the highest validation score, denoted \(L_H(t)\).  The harmfulness carrier is then
\begin{equation}
    h_t=\frac{\Delta^H_{L_H(t),t}}{\|\Delta^H_{L_H(t),t}\|_2}.
    \label{eq:harmfulness_carrier}
\end{equation}

The refusal carrier \(r_t\) is obtained from refusal-versus-compliance contrasts, following refusal-direction and COSMIC-style carrier extraction methods \cite{arditi2024refusal,siu2025cosmic}. The paired use of harmfulness and refusal carriers is motivated by evidence that LLMs can encode harmfulness and refusal as distinct latent concepts \cite{zhao2025harmfulnessrefusal}.  For each candidate token position \(p\) and layer \(\ell\), we compute
\begin{equation}
    \Delta^R_{p,\ell,t}
    =
    \mathbb{E}\!
    \left[a^{(\ell)}_{p}(x;M_t)\mid \mathrm{refusal}\right]
    -
    \mathbb{E}\!
    \left[a^{(\ell)}_{p}(x;M_t)\mid \mathrm{compliance}\right].
    \label{eq:refusal_carrier_delta}
\end{equation}
We search over candidate positions and layers and select the best admissible COSMIC-style carrier \cite{siu2025cosmic}, with admissibility determined by the refusal score, compliance side effect, and Kullback--Leibler (KL) divergence filters.  If \((p_R(t),L_R(t))\) is the selected position--layer pair, the refusal carrier is
\begin{equation}
    r_t=\frac{\Delta^R_{p_R(t),L_R(t),t}}{\|\Delta^R_{p_R(t),L_R(t),t}\|_2}.
    \label{eq:refusal_carrier}
\end{equation}
When a top-\(k\) local subspace is used, we write \(\mathcal{H}_t\) and \(\mathcal{R}_t\) for the orthonormal bases associated with harmfulness and refusal, respectively.  These are operational measurement objects: \(h_t\) is a hidden-state contrast direction for harmful inputs relative to benign inputs, and \(r_t\) is a hidden-state contrast direction for refusal behavior relative to compliance behavior.  Neither carrier is a manually specified word vector, and we do not equate \(h_t\) with the full harmfulness representation or \(r_t\) with the entire refusal mechanism.

\subsection{Harmfulness--refusal coupling}

We summarize H/R coupling using three normalized quantities.  The first is directional alignment,
\begin{equation}
    C_{\mathrm{cos}}(t)=\left|h_t^{\top}r_t\right|.
\end{equation}
For subspaces, let \(\theta_1,\ldots,\theta_k\) be the first \(k\) principal angles between \(\mathcal{H}_t\) and \(\mathcal{R}_t\), a standard way to measure overlap between linear subspaces \cite{bjorck1973angles}. We use the average squared canonical correlation,
\begin{equation}
    C_{\mathrm{sub}}(t)=\frac{1}{k}\sum_{i=1}^{k}\cos^2\theta_i,
\end{equation}
which is high when the two local subspaces overlap and low when their leading directions are nearly orthogonal.  Our primary direct representational coupling index is
\begin{equation}
    \mathrm{HRCI}_{\mathrm{repr}}(t)
    =\frac{1}{2}C_{\mathrm{cos}}(t)
     +\frac{1}{2}C_{\mathrm{sub}}(t).
    \label{eq:hrci_repr}
\end{equation}
The equal weighting is a fixed symmetry-based summary of two direct activation-space overlap measures. It was not optimized against ASR, refusal, or utility outcomes; we therefore interpret \hrcirepr{} descriptively and report component-wise and sensitivity analyses separately.  Because selected carrier layers are discrete and may be sensitive to direction-selection details, we report normalized layer separation,
\begin{equation}
    d_{\mathrm{layer}}(t)=\frac{|L_H(t)-L_R(t)|}{L-1},
    \label{eq:layer_sep}
\end{equation}
as a separate localization diagnostic.  For continuity with the initial trajectory summary, we also report layer co-localization,
\begin{equation}
    C_{\mathrm{layer}}(t)=\exp\left(-\frac{|L_H(t)-L_R(t)|}{\tau}\right),
    \qquad \tau=4.
\end{equation}
Here \(\tau\) is an e-folding layer scale rather than a decision threshold; in the 32-layer Mistral backbone, \(\tau=4\) is approximately one eighth of model depth.  The full weighted diagnostic is
\begin{equation}
    \mathrm{HRCI}_{\mathrm{full}}(t)
    =0.4C_{\mathrm{cos}}(t)
     +0.4C_{\mathrm{sub}}(t)
     +0.2C_{\mathrm{layer}}(t).
    \label{eq:hrci}
\end{equation}
For example, at the R2D2 reference checkpoint the source table gives \(C_{\mathrm{cos}}=0.0977\), \(C_{\mathrm{sub}}=0.0370\), and \(C_{\mathrm{layer}}=0.1738\); Eq.~\eqref{eq:hrci} yields \(\mathrm{HRCI}_{\mathrm{full}}=0.0886\), matching the continuity column in Table~\ref{tab:five_anchor}.  Both summaries are operational trajectory diagnostics, not ground-truth latent variables.

\subsection{Orthogonalized causal interventions and controls}

The causal matrix evaluates H-only, R-only, H+R, \(H_{\perp R}\), and \(R_{\perp H}\) interventions under ablation and steering.  To separate shared from residual components, we form
\begin{equation}
    h_{\perp r}=\frac{h_t-(h_t^{\top}r_t)r_t}{\|h_t-(h_t^{\top}r_t)r_t\|_2},
    \qquad
    r_{\perp h}=\frac{r_t-(r_t^{\top}h_t)h_t}{\|r_t-(r_t^{\top}h_t)h_t\|_2}.
    \label{eq:orthogonalized}
\end{equation}
These residual carriers are diagnostic controls.  Orthogonality in activation space does not imply mechanistic independence.

For a unit carrier \(u\), activation \(a\), and intervention strength \(\lambda\), we write projection ablation as
\begin{equation}
    T_{\mathrm{abl}}(a;u,\lambda)=a-\lambda(u^{\top}a)u,
    \label{eq:ablation}
\end{equation}
and additive steering as
\begin{equation}
    T_{\mathrm{steer}}(a;u,\lambda)=a+\lambda u.
    \label{eq:steering}
\end{equation}
Equation~\eqref{eq:steering} is the normalized operator form used for exposition; any implementation that applies an additional norm-scaling constant should report that constant explicitly rather than absorbing it into the interpretation of \(\lambda\).  The full Mistral H/R causal sweep covers SFT and R2D2, five anchors, ablation/steering families, H/R modes, and lambda strengths \(\{-4,-2,-1,-0.5,0,0.5,1,2,4\}\) where applicable.

Controls are matched same-layer random unit directions, wrong-layer carriers, and wrong-position carriers. These controls address several carrier-selection artifacts, but they are not raw activation-delta norm-matched controls. We therefore phrase the causal result as fixed-protocol sensitivity relative to matched unit-direction controls, not as evidence that all norm or scale artifacts have been ruled out.

\subsection{Behavioral metrics and diagnostic effects}

For a harmful evaluation set \(\mathcal{D}_{\mathrm{harm}}\), attack success rate is
\begin{equation}
    \mathrm{ASR}(M)=\frac{1}{|\mathcal{D}_{\mathrm{harm}}|}
    \sum_{x\in\mathcal{D}_{\mathrm{harm}}}\ind\{\mathrm{success}(M,x)=1\}.
    \label{eq:asr}
\end{equation}
For StrongREJECT, where \(s_{\mathrm{SR}}(M,x)\in[0,1]\) scores harmful usefulness, we report
\begin{equation}
    \mathrm{SR}(M)=\frac{1}{|\mathcal{D}_{\mathrm{SR}}|}
    \sum_{x\in\mathcal{D}_{\mathrm{SR}}}s_{\mathrm{SR}}(M,x).
    \label{eq:strongreject}
\end{equation}
XSTest any-refusal measures whether safe prompts receive full or partial refusals \cite{rottger2024xstest}.  It is defined as
\begin{equation}
    \mathrm{XR}(M)=\frac{1}{|\mathcal{D}_{\mathrm{safe}}|}
    \sum_{x\in\mathcal{D}_{\mathrm{safe}}}\ind\{R_{\mathrm{full}}(M,x)\vee R_{\mathrm{partial}}(M,x)\}.
    \label{eq:xstest}
\end{equation}
For benign utility annotations \(u(M,x)\in\{0,1,2\}\), we separately report
\begin{align}
    U_{\mathrm{strict}}(M)&=\frac{1}{n}\sum_x\ind\{u(M,x)=2\}, \\
    U_{\mathrm{lenient}}(M)&=\frac{1}{n}\sum_x\ind\{u(M,x)\ge 1\}, \\
    U_{\mathrm{mean}}(M)&=\frac{1}{n}\sum_x u(M,x), \\
    D_{\mathrm{deg}}(M)&=\frac{1}{n}\sum_x\ind\{\mathrm{degenerate}(M,x)=1\}.
    \label{eq:utility}
\end{align}
This three-track utility reporting is necessary because reduced refusal on safe prompts does not by itself imply useful benign behavior.

For any outcome \(Y\), intervention mode \(m\), and baseline mode \(b\), the raw effect is
\begin{equation}
    \Delta_Y(m,t)=Y(M_t^{m})-Y(M_t^{b}).
\end{equation}
Given a matched control set \(\mathcal{C}(m,t)\), the control-normalized effect is
\begin{equation}
    \Delta_Y^{\mathrm{ctrl}}(m,t)=\Delta_Y(m,t)-
    \frac{1}{|\mathcal{C}(m,t)|}\sum_{c\in\mathcal{C}(m,t)}\Delta_Y(c,t).
    \label{eq:ctrl_delta}
\end{equation}
All control-normalized claims in this paper use the matched unit-direction controls described above.

\subsection{Four-quadrant behavior construction and sparse transfer}

We construct a four-quadrant H/R audit set with explicit labels for harmful input \(H_{\mathrm{in}}\), refusal output \(R_{\mathrm{out}}\), benign utility \(U_{\mathrm{out}}\), and degeneration \(D_{\mathrm{out}}\).  The diagnostic quadrants are
\begin{equation}
    \mathcal{Q}_{h,r}=\{(x,y):H_{\mathrm{in}}(x)=h,\ R_{\mathrm{out}}(y)=r\},
    \qquad h,r\in\{0,1\}.
    \label{eq:quadrants}
\end{equation}
They correspond to the four input--output combinations \(H=1,R=1\), \(H=1,R=0\), \(H=0,R=0\), and \(H=0,R=1\).  Benign non-refusal is not automatically counted as helpful: \(D_{\mathrm{out}}\) and \(U_{\mathrm{out}}\) are reported separately.  The construction samples 500 rows per canonical quadrant from available fixed-protocol outputs, with an additional benign-degenerate audit category. This artifact supports diagnostic analysis of the current protocol; it is not presented as a new general-purpose benchmark without further human validation.

For sparse transfer, let \(\mathcal{P}_{s,A}\) be prompts generated by attack family \(A\) at source checkpoint \(s\).  Transfer ASR at target checkpoint \(t\) is
\begin{equation}
    \mathrm{TASR}_{A}(s\rightarrow t)=
    \frac{1}{|\mathcal{P}_{s,A}|}\sum_{p\in\mathcal{P}_{s,A}}
    \ind\{\mathrm{success}(M_t,p)=1\}.
    \label{eq:tasr}
\end{equation}
We compare geometry and behavior predictors using Spearman correlation,
\begin{equation}
    \rho_S(X,Y)=\mathrm{corr}(\mathrm{rank}(X),\mathrm{rank}(Y)).
    \label{eq:spearman}
\end{equation}
These correlations are descriptive diagnostics for sparse transfer, not proof for model selection.

\subsection{Statistical protocol and reporting scope}

The trajectory checkpoints are ordered states along two training trajectories rather than independent random samples. We therefore use signed correlations, block-bootstrap intervals, and control-normalized contrasts as descriptive trajectory diagnostics, not as independent and identically distributed (iid) population estimates.

A provenance audit found that an earlier all-anchor HRCI table mixed refusal-carrier artifacts produced under incompatible extraction settings. We therefore use a quality-controlled all-anchor table in which refusal-side carriers were extracted under one protocol and aligned with verified H-side artifacts. Inclusion required exact agreement with the five-anchor reference set across 80 anchor--field consistency checks, with 0 failures. Sparse-transfer HRCI summaries are computed by matching the completed GCG/AutoDAN transfer rows to this quality-controlled table and are reported only as small-\(n\) HRCI-vs-drift diagnostics.

Causal tables first use mode-level summaries over the completed intervention matrix and then show representative rows. When a result is selected from top control-normalized rows, the selection rule and matched controls are reported in the source tables. No claim in this paper depends on multiple training seeds or on deployment-time generalization.

\section{Results}
\label{sec:results}

We organize the main results around five research questions (RQs).

\subsection{RQ1: Do H and R measure distinct safety functions?}
\label{sec:aligned}

The aligned-anchor calibration asks whether the H/R protocol separates harmfulness-sensitive and refusal-sensitive directions in models that already have mature instruction-following behavior. Table~\ref{tab:aligned} reports the main pattern. In Llama-3.1-8B-Instruct \cite{grattafiori2024llama3}, H and R are selected at different layers, and their leading principal angles are close to orthogonal. R-only and \(R_{\perp H}\) ablations reopen HarmBench ASR sharply, while H-only ablation does not. Qwen2.5-7B-Instruct \cite{yang2024qwen25} shows the same qualitative refusal-side bottleneck, though the H-side effect is less clean. The calibration supports a dual-carrier measurement protocol; it does not prove two independent neural pathways.

\begin{table}[!htbp]
\centering
\caption{Aligned-anchor H/R calibration. ASR values are HarmBench attack success rates under the aligned-anchor causal protocol and are used for within-protocol intervention comparison, not as a complete safety evaluation of the aligned models. The pattern exposes refusal-side reopening without proving independent H/R mechanisms.}
\label{tab:aligned}
\small
\setlength{\tabcolsep}{4pt}
\begin{tabular}{@{}lcc@{}}
\toprule
Metric & Llama-3.1-8B & Qwen2.5-7B \\
\midrule
H layer & 31 & 15 \\
R layer & 12 & 20 \\
Principal angles & 86.1, 88.7, 89.4 & 82.3, 88.1, 89.7 \\
Baseline ASR & 0.1675 & 0.2625 \\
H-only ASR & 0.1400 & 0.2925 \\
R-only ASR & 0.6825 & 0.4050 \\
\(R_{\perp H}\) ASR & 0.6575 & 0.4250 \\
\bottomrule
\end{tabular}
\end{table}

This calibration is necessary for interpreting the Mistral trajectory. If the same protocol did not recover a refusal-side bottleneck in aligned instruction models, a mixed result on base-to-R2D2 checkpoints would be hard to interpret. Instead, the aligned anchors show that the protocol can detect refusal-side causal reopening under a fixed intervention protocol.

\subsection{RQ2: How does R2D2 change H/R coupling?}
\label{sec:trajectory}

\begin{figure*}[!t]
\centering
\includegraphics[width=0.98\textwidth]{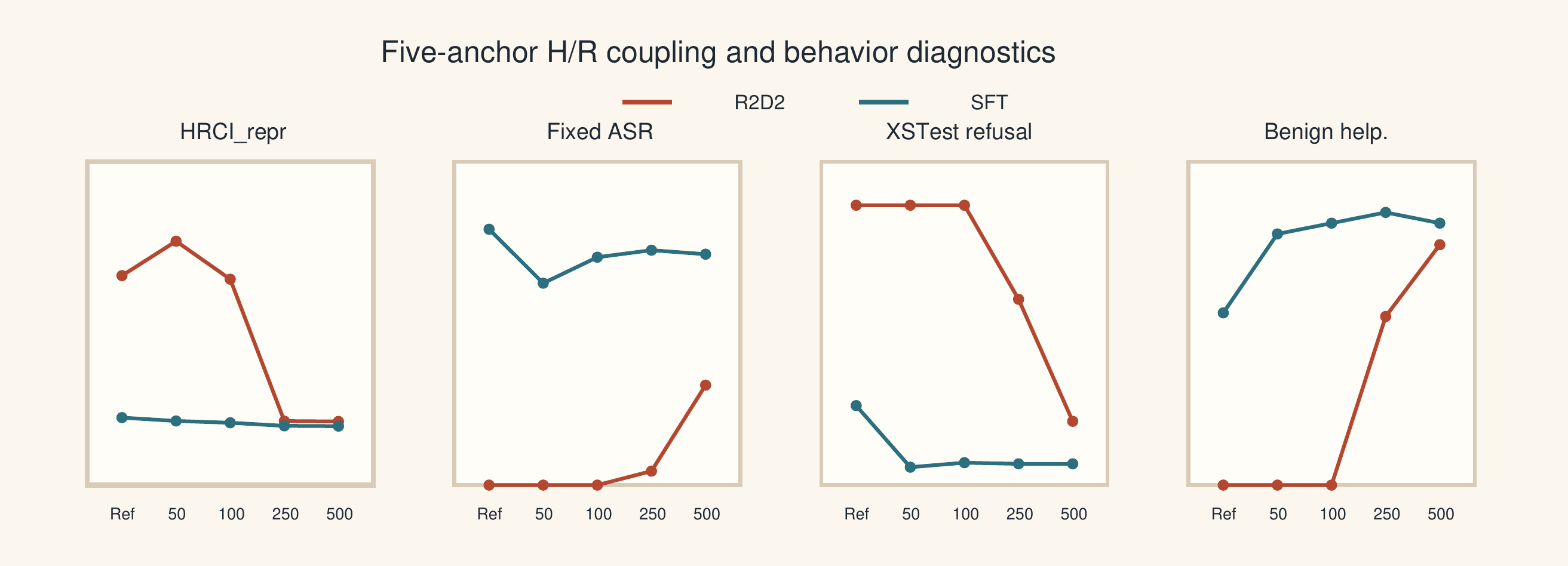}
\caption{Five-anchor H/R coupling and behavior diagnostics for Mistral SFT and R2D2. The panels align the primary \hrcirepr{} summary, fixed-source HarmBench ASR, XSTest any-refusal, and benign helpfulness on the same anchor sequence. Early R2D2 is relatively more coupled, robust, and utility-collapsed; later R2D2 is lower-coupling, partly useful, and partly fail-open.}
\label{fig:trajectory_diagnostic}
\end{figure*}

Unlike the prior refusal-geometry trajectory analysis \cite{lan2026refusalgeometry}, this section treats the refusal carrier as one component of a dual H/R system and asks how its coupling with harmfulness changes over time. Table~\ref{tab:five_anchor} reports the five-anchor Mistral trajectory, with the original full HRCI retained as a continuity column. The strongest pattern is the R2D2 transition between steps 100 and 250. Direct representational coupling is relatively higher at the early R2D2 anchors: \hrcirepr{} is 0.0784 at step 50, while fixed-source ASR remains 0 and XSTest refusal is saturated at 1.00. Benign helpfulness is also 0, so this early state is robust-looking but not useful. By step 500, \hrcirepr{} has fallen to 0.0205, a 73.9\% drop from step 50. At the same checkpoint, fixed-source ASR has reopened to 0.2500, XSTest refusal has fallen to 0.2280, and benign helpfulness has recovered to 1.1167 on a 0--2 scale. The full HRCI continuity value shows the same direction, dropping from 0.0898 to 0.0165. Figure~\ref{fig:trajectory_diagnostic} shows the same transition on a shared anchor axis.

\begin{table*}[!t]
\centering
\caption{Five-anchor H/R coupling and behavior summary for Mistral SFT/R2D2. Helpfulness is the mean 0--2 benign-utility score on the 60-prompt continuity set. \hrcirepr{} is the primary direct-coupling metric in Eq.~\eqref{eq:hrci_repr}; full HRCI is retained as the continuity parameterization in Eq.~\eqref{eq:hrci}.}
\label{tab:five_anchor}
\small
\setlength{\tabcolsep}{4pt}
\begin{tabular}{llcccccc}
\toprule
Regime & Anchor & \hrcirepr{} & Full HRCI & Fixed ASR & XSTest refusal & StrongREJECT & Benign help. \\
\midrule
\multirow{5}{*}{R2D2}
& Reference & 0.0674 & 0.0886 & 0.0000 & 1.0000 & 0.1262 & 0.0000 \\
& Step 50   & 0.0784 & 0.0898 & 0.0000 & 1.0000 & 0.1220 & 0.0000 \\
& Step 100  & 0.0662 & 0.0877 & 0.0000 & 1.0000 & 0.1707 & 0.0000 \\
& Step 250  & 0.0206 & 0.0166 & 0.0350 & 0.6640 & 0.2484 & 0.7833 \\
& Step 500  & 0.0205 & 0.0165 & 0.2500 & 0.2280 & 0.2638 & 1.1167 \\
\midrule
\multirow{5}{*}{SFT}
& Reference & 0.0217 & 0.0521 & 0.6400 & 0.2840 & 0.2172 & 0.8000 \\
& Step 50   & 0.0206 & 0.0166 & 0.5050 & 0.0640 & 0.3216 & 1.1667 \\
& Step 100  & 0.0200 & 0.0161 & 0.5700 & 0.0800 & 0.3952 & 1.2167 \\
& Step 250  & 0.0191 & 0.0153 & 0.5875 & 0.0760 & 0.4530 & 1.2667 \\
& Step 500  & 0.0190 & 0.0153 & 0.5775 & 0.0760 & 0.4645 & 1.2167 \\
\bottomrule
\end{tabular}
\end{table*}

\subsection{RQ3: Is lower coupling a general signature of safety?}

The SFT trajectory rules out a simple monotonic interpretation. SFT reaches full-HRCI values close to late R2D2 by step 50, but it remains high-ASR throughout the five-anchor window. Its direct-coupling change is also much smaller: \hrcirepr{} moves from 0.0217 at reference to 0.0190 at step 500. Thus low coupling is not a safety score. The useful signal is the regime transition: in R2D2, direct coupling changes jointly with robustness, over-refusal, and benign utility; in SFT, low coupling coexists with persistent vulnerability.

\subsection{RQ4: Does the conclusion depend on a few anchors or selected intervention rows?}

The dense behavior matrix expands beyond the five anchors to 100 checkpoints per regime. Coverage is complete for XSTest and benign utility, with a dense StrongREJECT \cite{souly2024strongreject} subset of 28 anchors per regime. Missing behavior shards are zero. The quality-controlled all-anchor HRCI table matches the five-anchor reference checks with 0/80 consistency failures. Using the primary \hrcirepr{} metric, R2D2 shows the expected regime-dependent pattern (Table~\ref{tab:dense_hrci}). SFT is the contrast: \hrcirepr{} changes weakly from first to last, and its XSTest/refusal/utility correlations are weak or unstable. Dense all-anchor results therefore support a fixed-protocol R2D2 diagnostic, not a general safety score. They also refine the five-anchor story: R2D2 direct coupling is a plateau-to-low transition around steps 105--110, not a monotonic decline from the earliest checkpoint.

\begin{table}[!htbp]
\centering
\caption{Dense all-anchor signed Spearman associations for R2D2 using the primary \hrcirepr{} metric. Moving-block bootstrap (MBB) intervals reflect checkpoint ordering and are descriptive rather than iid confidence intervals.}
\label{tab:dense_hrci}
\small
\setlength{\tabcolsep}{4pt}
\begin{tabular}{lccc}
\toprule
Outcome & \(\rho\) & \(n\) & MBB 95\% interval \\
\midrule
Fixed-source ASR & -0.646 & 100 & [-0.901, -0.256] \\
XSTest any-refusal & 0.659 & 100 & [0.273, 0.917] \\
Benign refusal rate & 0.725 & 100 & [0.376, 0.935] \\
Benign helpfulness & -0.661 & 100 & [-0.898, -0.293] \\
StrongREJECT score & -0.640 & 28 & [-0.957, -0.139] \\
\bottomrule
\end{tabular}
\end{table}

The H/R causal-sweep matrix gives the same scoped message. It covers SFT and R2D2, ablation and steering, five anchors, H-only, R-only, H+R, \(H_{\perp R}\), \(R_{\perp H}\), and matched unit-direction controls. Mode-level summaries show that the strongest control-normalized fixed-source reopening is concentrated in SFT R-side rows and selected steering rows. Table~\ref{tab:wp6} gives representative rows, but the interpretation comes from the matrix: many R2D2 ablation modes do not exceed control maxima, and matched controls can have comparable effects.

\begin{table*}[!t]
\centering
\caption{Representative control-normalized H/R causal-sweep rows. \(\Delta\)control is the target fixed-source ASR minus the mean of matched same-layer random, wrong-layer, and wrong-position unit-direction controls. These rows motivate intervention-level follow-up; they are not a full proof of independent pathways.}
\label{tab:wp6}
\small
\begin{tabular}{llcccc}
\toprule
Run family & Selected row & Anchor & Operation & Target ASR & \(\Delta\)control \\
\midrule
R2D2 ablation & H-only & Reference & ablate 0.25 & 0.1437 & 0.0325 \\
R2D2 ablation & \(H_{\perp R}\) & Reference & ablate 0.25 & 0.1437 & 0.0325 \\
SFT ablation & \(R_{\perp H}\) & Step 500 & ablate 0.5 & 0.3344 & 0.0719 \\
SFT ablation & R-only & Step 250 & ablate 0.5 & 0.3406 & 0.0675 \\
R2D2 steering & \(H_{\perp R}\) & Step 500 & add -4 & 0.1406 & 0.0544 \\
R2D2 steering & H-only & Step 500 & add -4 & 0.1375 & 0.0513 \\
SFT steering & \(R_{\perp H}\) & Step 250 & add 2 & 0.3906 & 0.1369 \\
SFT steering & R-only & Step 250 & add 4 & 0.3906 & 0.1362 \\
\bottomrule
\end{tabular}
\end{table*}

The 43-row intervention-level follow-up checks whether reopening corresponds to harmful usefulness or generic disruption. Across the 8 selected target rows, mean fixed-source ASR is 0.253, compared with 0.207 across 30 matched controls. However, selected targets also have high empty-or-nonsensical rates on StrongREJECT (mean 0.833) and low benign helpfulness (mean 0.288). The SFT steering targets are especially cautionary: they reach ASR 0.3906, but their mean StrongREJECT score is only 0.024 and benign helpfulness is 0. This follow-up therefore supports fixed-protocol sensitivity, but it does not support a clean harmful-usefulness reopening claim.

\subsection{RQ5: What do sparse-transfer HRCI-vs-drift diagnostics show?}
\label{sec:transfer}

The sparse-transfer block evaluates GCG and AutoDAN attacks generated at source checkpoints and transferred to target checkpoints. For SFT, source generation covers 120/120 shards and transfer covers 240/240 off-diagonal rows, with zero failed rows. The R2D2 counterpart uses the same fixed-protocol diagnostic structure. We match these transfer rows to the quality-controlled all-anchor H/R table and compare \hrcirepr{} predictors against behavior-state and refusal-drift predictors. For the drift-only baseline, target refusal-score drift denotes \texttt{target\_refusal\_score\_drift\_ref}: the target checkpoint's refusal-carrier score drift relative to the regime-specific reference checkpoint, as recorded in the sparse-transfer source table. It is distinct from the target--source refusal-drift gap, which measures the difference between target and source drift values and is not the Table~\ref{tab:transfer} drift-only predictor. This provides a quality-controlled HRCI-transfer linkage, but the resulting table is still a sparse 12-row-per-regime diagnostic rather than a black-box or adaptive validation suite.

\begin{table*}[!t]
\centering
\caption{Sparse-transfer comparison between target \hrcirepr{}, fixed-source ASR, refusal-score drift, and target--source \hrcirepr{} gap. Values are signed Spearman correlations with transfer ASR over pooled GCG and AutoDAN transfer rows. Refusal-score drift is measured relative to the regime-specific reference checkpoint using the refusal-carrier score recorded in the sparse-transfer source table. The result is descriptive and fixed-protocol; it is not full black-box validation, adaptive robustness evidence, or a model-selection proof.}
\label{tab:transfer}
\small
\begin{tabular}{llcc}
\toprule
Regime & Predictor & Family & \(\rho\) \\
\midrule
R2D2 & Target \hrcirepr{} & geometry & -0.965 \\
R2D2 & Target fixed-source ASR & behavior state & 0.965 \\
R2D2 & Target refusal-score drift & drift only & 0.965 \\
R2D2 & Target--source \hrcirepr{} gap & geometry gap & -0.735 \\
\midrule
SFT & Target \hrcirepr{} & geometry & -0.591 \\
SFT & Target refusal-score drift & drift only & -0.591 \\
SFT & Target fixed-source ASR & behavior state & 0.562 \\
SFT & Target--source \hrcirepr{} gap & geometry gap & -0.424 \\
\bottomrule
\end{tabular}
\end{table*}

The sparse-transfer comparison supports the same caution as the dense table. In R2D2, target \hrcirepr{} tracks transfer ASR strongly, but it is tied with target fixed-source ASR and target refusal-score drift in this sparse summary. In SFT, target \hrcirepr{} and refusal-score drift are weaker and again similar in magnitude. Thus the transfer block checks the all-anchor HRCI linkage and shows descriptive HRCI--transfer alignment, but it does not show added predictive value beyond visible behavior-state or refusal-drift predictors.

\section{Discussion}

\paragraph{Principal finding.}
The main result is not that a lower HRCI value is better, nor that harmfulness and refusal are independent mechanisms. The result is that R2D2 moves through two H/R coupling regimes that have different behavior profiles. The early regime is robust on fixed-source attacks but unusable on benign prompts; the later regime recovers part of benign utility while reopening attacks. This explains why refusal drift alone is incomplete: the relevant object is the relation between harmfulness recognition and refusal control, not only the position of a refusal carrier.

\paragraph{Mechanistic interpretation.}
The aligned anchors show that refusal-side carriers can be causal bottlenecks in already aligned instruction models. The Mistral five-anchor trajectory shows a different phenomenon: adversarial fine-tuning first produces a relatively more coupled safety state and then relaxes direct representational coupling as utility returns. One plausible interpretation is that early R2D2 suppresses attack success by binding harmfulness recognition tightly to refusal activation, which also collapses benign utility. Later R2D2 recovers more benign behavior, suggesting a more differentiated refusal controller; that differentiation is not automatically safer, because it coincides with partial attack reopening. SFT is the guardrail against overinterpretation: it also has low coupling, but remains high-ASR. The evidence therefore supports a training-dynamics explanation, not a claim that low coupling is intrinsically good.

The evidence does not identify a single component as the whole mechanism. Directional alignment and subspace overlap define the primary \hrcirepr{} diagnostic, while layer separation is reported separately as localization evidence. A natural next analysis is to expand sparse transfer beyond the present GCG/AutoDAN block and align the coupling transition with loss traces, attack-pool refreshes, and carrier-selection changes.

The causal interventions establish functional sensitivity under the selected carriers, but they do not identify all safety-relevant circuits or rule out distributed alternatives. Multi-directional and fail-closed refusal mechanisms may require redundant or distributional safety features; \hrcirepr{} does not attempt to measure such redundancy, and instead tracks coupling between two selected H/R carrier families under a fixed protocol.

\paragraph{Relation to nearby work.}
Single-direction refusal work shows that refusal can be controlled through activation directions \cite{arditi2024refusal}. COSMIC and concept-cone analyses make that picture less template-dependent and more subspace-aware \cite{siu2025cosmic,wollschlager2025geometry}. Harmfulness--refusal separation work shows why refusal should not be treated as the only safety concept \cite{zhao2025harmfulnessrefusal}. Our contribution is to put these ideas into a training trajectory. Compared with the related refusal-geometry manuscript \cite{lan2026refusalgeometry}, this paper changes both the explanatory variable and the claim. The prior study treated refusal control as the primary internal object and concluded that R2D2 reorganizes a low-dimensional, utility-coupled refusal carrier along a robustness--utility frontier. Here, refusal control is only one side of the measurement. The central object is the relation between harmfulness recognition and refusal control, which we test through aligned anchors, coupling diagnostics, H/R interventions, dense trajectory checks, and sparse-transfer analyses. Thus the claim is not that R2D2 merely relocates a refusal carrier, but that it passes through distinct harmfulness--refusal coupling regimes whose behavior profiles differ in robustness, over-refusal, and benign utility. Compared with defense-oriented methods such as ReFAT, latent adversarial training, and projection-constrained tuning \cite{yu2025refat,casper2024latentAdversarial,sheshadri2024targetedLat,du2025procon}, the aim here is diagnosis rather than a new training algorithm.

\paragraph{Implications for training-time monitoring.}
The practical use of H/R coupling is not standalone deployment prediction. A more defensible use is joint monitoring during training. In the current evidence chain, the five-anchor \hrcirepr{} summary, fixed-source ASR, XSTest refusal, StrongREJECT, and benign utility together distinguish at least two failure states: a model that is robust because it refuses too much, and a model that becomes useful again while reopening attack success. The R2D2 trajectory contains both states. That makes H/R coupling useful as a candidate warning signal when interpreted with behavior readouts, not as an isolated scalar objective.

\paragraph{Alternative explanations and claim boundary.}
Several alternatives remain. The measured carriers may be affected by layer-selection noise, checkpoint autocorrelation, norm and scale artifacts, or degeneration under intervention. The causal matrix uses matched unit-direction controls, not controls matched to raw activation-delta norms. Sparse transfer covers GCG and AutoDAN under a fixed protocol, not black-box, multi-turn, or adaptive attacker-model search. The sparse-transfer HRCI comparison is also small-\(n\) and partly collinear with behavior-state predictors. These boundaries are why we describe the evidence as fixed-protocol causal sensitivity and regime-dependent monitoring evidence, not as proof of independent pathways, fail-closed redundancy, or deployment-grade prediction.

\section{Limitations}

The main trajectory uses one backbone, Mistral-7B-v0.1, under matched SFT and R2D2 regimes. The aligned Llama and Qwen anchors calibrate the protocol, but they do not provide a full multi-backbone trajectory atlas. Future work should repeat the same trajectory analysis on Llama, Qwen, Gemma-family, and other alignment regimes before treating the observed transition as a general training law.

HRCI is an operational diagnostic, not a mechanistic ground truth. The primary \hrcirepr{} uses fixed equal weights over directional and subspace overlap, while the full HRCI parameterization is retained only for continuity with earlier trajectory summaries. The dense all-anchor correlations and sparse-transfer joins are descriptive and fixed-protocol, not deployment predictors. Future work should report component-wise HRCI, weight sensitivity, learned combinations, and Pareto-style alternatives across additional model families.

The causal evidence is also protocol-limited. The H/R causal-sweep controls are matched same-layer random, wrong-layer, and wrong-position unit-direction controls. We do not currently include controls matched to raw activation-delta norms, so we avoid claims that all norm or scale artifacts are ruled out. The intervention-level StrongREJECT and benign-utility follow-up covers selected top rows and matched controls, not the full H/R causal-sweep matrix. Future work should combine raw norm-matched controls with full intervention-level behavior and utility measurements.

Finally, the evaluation suite is intentionally bounded. Benign utility is measured with a 60-prompt continuity set rather than a 500--1000 prompt utility benchmark. Sparse transfer covers GCG and AutoDAN under a fixed protocol; Prompt Automatic Iterative Refinement (PAIR)- and Tree of Attacks with Pruning (TAP)-style attacker-model search and broader black-box suites remain outside the evidence chain \cite{chao2023pair,mehrotra2024tap}. The four-quadrant construction is an audit artifact and should be human-validated before being treated as a benchmark. Future work should scale the benign-utility benchmark, add broader adaptive and black-box attack suites, and validate the four-quadrant construction independently.

\section{Conclusion}

Dynamic adversarial fine-tuning changes not only how much a model refuses, but how harmfulness recognition and refusal control are coupled. In the five-anchor R2D2 trajectory, that coupling-regime transition explains a robustness--over-refusal--utility pattern that refusal drift alone does not capture. The claim is deliberately scoped: direct H/R coupling is a fixed-protocol training diagnostic to be read with behavior metrics, not a proof of independent pathways or a deployment-ready safety predictor.

\section{Safety and Ethics}

This work evaluates harmful-request refusal and jailbreak robustness. The experiments are reported through aggregate metrics, intervention summaries, and diagnostic tables rather than by reproducing harmful prompts, optimized suffixes, or harmful completions. The goal is not to provide a jailbreak method, but to understand when adversarial fine-tuning creates robust refusal, broad over-refusal, benign utility collapse, or fail-open behavior. Release of code or artifacts should preserve this separation by withholding raw harmful generations and adversarial strings where needed.

\section*{Acknowledgements}
The authors acknowledge the public benchmark, model, and software resources cited in this work.

\section*{Competing interests}
The authors declare that they have no competing interests or financial conflicts to disclose.

\section*{Data and Code Availability}
The supplementary artifact package contains the aggregate source tables, figure-source files, run manifests, configuration and compute-provenance summaries, and analysis and visualization scripts needed to reproduce the paper-level tables and figures. A stable mapping from each analysis component to its supplementary artifact identifier is provided in the supplementary manifest files rather than as a main-text appendix table.

Upon acceptance, we will deposit a public archival version of the artifact package in a long-term repository and add the resulting DOI to the final article. The archival release will include versioned aggregate data, figure sources, analysis and plotting code, configuration summaries, checksums, and provenance records. Third-party model weights and benchmark resources will remain subject to their original licenses and will not be redistributed.

To limit dual-use risk, the public archive will not include optimized adversarial strings or generated harmful completions. Raw harmful prompts will be represented only through references to public benchmark sources or through redacted records where appropriate. The released materials are intended to support verification of the reported aggregate analyses and regeneration of the paper's figures and tables, rather than unrestricted reproduction of harmful attack content.

\bibliographystyle{unsrt}
\bibliography{ref}

\end{document}